\documentclass[letterpaper,12pt]{article}

\usepackage{graphics,graphicx,apjfonts}

\makeatletter
\renewcommand{\section}{\@startsection%
{section}{1}{0mm}{-\baselineskip}%
{0.5\baselineskip}{\normalfont\Large\bfseries}}%
\makeatother

\begin{document}

\noindent
\null
\begin{center}
{\bf \large A JavaScript Passive Evolution Calculator}
\end{center}
When there is no new star formation, galaxies evolve ``passively'', i.e.,
they redden and fade as stars turn off of the
main sequence. In principle,
this evolution can be measured by comparing the mass-to-light
ratios and colors
of distant passively evolving galaxies to those of nearby ones.
The measured evolution can then be compared to predictions from stellar
population syntesis models, to determine ages and other parameters.

This technique has been applied successfully to determine the
ages of early-type galaxies
in the field and clusters, out to redshifts $z\sim 1.25$ (e.g.,
van Dokkum \& Franx 1996; Treu et al.\ 2002;
van der Wel et al.\ 2004).
However, there is growing evidence that
the number of early-type galaxies is not stable with time,
but decreases with redshift (e.g., Dressler et al.\ 1997;
van Dokkum et al.\ 2000; Bell et al.\ 2004). The implication
of such complex evolution is that the sample of early-type
galaxies at high redshift is a biased subset of the sample
at $z=0$, containing only the oldest progenitors of today's
early-types. This ``progenitor bias'' leads us to overestimate
the ages of today's early-type galaxies.

In van Dokkum \& Franx (2001), we quantified the effects of
progenitor bias on the observed luminosity and color evolution
of early-type galaxies, and developed a simple model which can
provide the
bias-corrected ages of early-type galaxies for an assumed
rate of morphological evolution. Here we
present a JavaScript calculator for predicting the luminosity
and color evolution of passively evolving galaxies, based
on the van Dokkum \& Franx (2001) parameterization.
\vspace{0.3cm}\\
{\bf Simple evolution}
\vspace{0.2cm}\\
In its simplest form, the calculator provides the evolution of
a stellar population with a given luminosity-weighted formation
redshift (``Default 1''). Figure 1 shows examples of the output, for
a stellar population formed at $z=6$.

\begin{figure}[h]
\includegraphics[scale=1,bb=53 427 510 636]{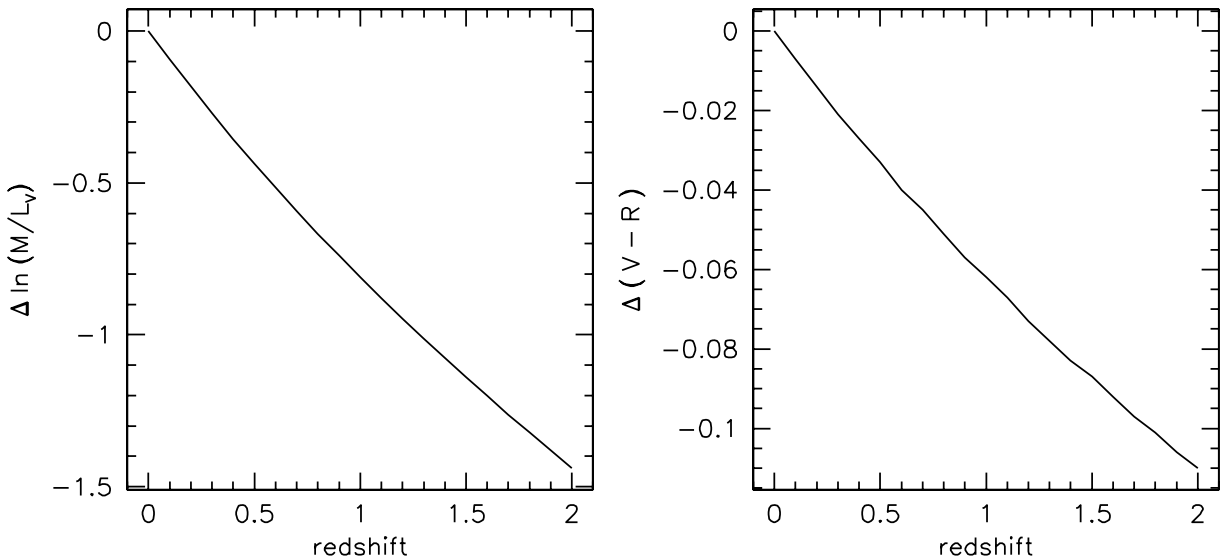}
\caption{\small Example output for a single stellar population formed at
$z=6$.}
\end{figure}

\noindent
{\bf Complex evolution}
\vspace{0.2cm}\\
The calculator can also incorporate morphological evolution (i.e.,
progenitor bias) and complex star formation histories of individual
galaxies. 
The calculator provides
the luminosity evolution, scatter in luminosity, color evolution,
scatter in color, the fraction of today's early-type galaxies 
formed at a given redshift, and the luminosity-weighted stellar
formation redshift of the early-types at that redshift.
The example below shows some of the output for ``Default 3'',
a model with strong morphological evolution.

\begin{figure}[h]
\includegraphics[scale=0.75,bb=-10 197 511 637]{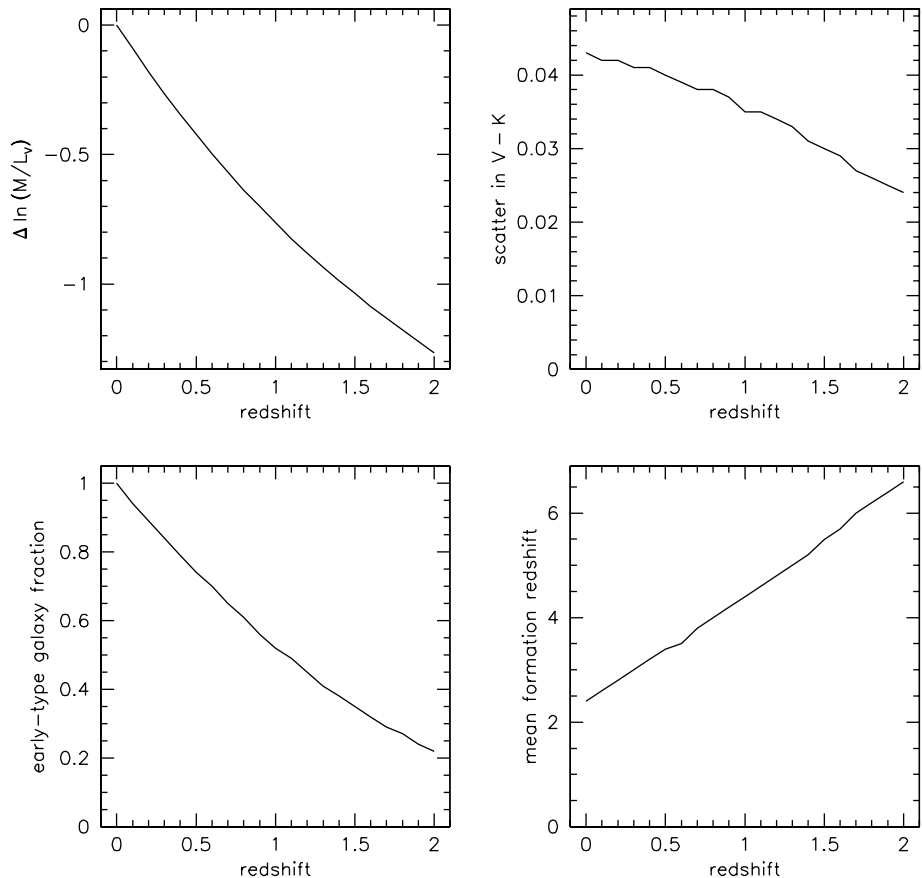}
\caption{\small Model with strong morphological evolution. Although only
$\sim 20$\,\% of present-day early-types were in place by $z=2$, the
luminosity evolution is very similar to that in Fig.\ 1, and the scatter
in $V-K$ color is small at all redshifts. These counter-intuitive trends
are explained in van Dokkum \& Franx (2001).}
\end{figure}

\noindent
Web location: {\bf www.astro.yale.edu/dokkum/evocalc/}\\
Please refer to van Dokkum \& Franx (2001) if you use this calculator.
\vspace{0.5cm}\\
{\small
Bell, E., et al.\ 2004, ApJ, 608, 752\\
Dressler, A., et al.\ 1997, ApJ, 490, 577\\
Treu, T., Stiavelli, M., Casertano, S., M/o{}ller, P., \& Bertin, G.
2002, ApJ, 564, L13\\
van der Wel, A., Franx, M., van Dokkum, P. G., Rix, H.-W. 2004, ApJ,
601, L5\\
van Dokkum, P. G., \& Franx, M. 1996, MNRAS, 281, 985\\
van Dokkum, P. G., Franx, M., Fabricant, D., Illingworth, G. D.,
\& Kelson, D. D. 2000, ApJ, 541, 95\\
van Dokkum, P. G., \& Franx, M. 2001, ApJ, 553, 90

\end{document}